\documentstyle[12pt, psfig]{article}
\begin{document}
\noindent {\Large \bf Liquid  Solid Transition of Hard Spheres Under Gravity}
\vskip 1.0 true cm
\centerline{ \bf Paul V. Quinn and Daniel C. Hong}
\vskip 0.2 true cm
\centerline{ Physics, Lewis Laboratory,
Lehigh University, Bethlehem, Pennsylvania 18015}
\vskip 0.3 true cm

\begin{abstract}
We investigate the liquid-solid transition of
two dimensional hard spheres in the presence of gravity.
We determine the transition temperature and the
fraction of particles in the solid regime as a function of temperature
via Even-Driven molecular dynamics simulations and compare them with
the theoretical predictions.  We then
examine the
configurational statistics of a vibrating bed 
from the view point of the liquid-solid transition by explicitly
determining the transition temperature and the effective
temperature, $T$, of the bed, and present a relation between $T$
and the vibration strength. 
\end{abstract}
\vskip 0.3 true cm
\noindent P.A.C.S: 05.20-y, 64.70Dv, 05.20.Dd, 51.10+y. 

\vskip 1.0 true cm
The hard sphere model has been quite successful in explaining the macroscopic
properties of dense fluids, or gases
from the microscopic point of view [1].
At the molecular level, the potential energy of the hard spheres
due to gravity is small in comparison to the thermal fluctuations
and it has usually been ignored. However, 
when the mass of the
constituent particle is macroscopic in quantity, as in the case of granular
materials [2], gravity cannot be ignored. 
The purpose of this Letter is to demonstrate the
existence of a gravity induced liquid-solid phase transition of hard spheres. 
This transition is an intrinsic transition associated with
any system where
the excluded volume interaction is dominant.  Such a
system cannot be compressed indefinitely, and must exhibit a coherent
low energy state.
In the hard sphere system, gravity introduces a potential energy,
and each available site is associated with an energy state.  Then,
the formation of a solid at the bottom below the transition 
point is nothing but a massive occupation of the
low energy state at the low temperature, which is the Fermi gas 
in metals,
the Bose condensate in the two dimensional quantum Hall systems [3],
the energy
storage mechanism into a single state for biological systems [4], and a
mechanism to produce coherent light in the context of lasers [5], and
the liquid-solid transition in a hard sphere system under gravity, which is the
subject of the current work. 
We will determine via Even-Driven Molecular Dynamics
simulation the transition point and the thickness of
the boundary layers as a function of external parameters, and make a careful
comparison with the theoretical predictions [6]. 
Next, a new and nontrivial
by-product of our investigation is to view the configurational
statistics of the vibrating bed [7,8] from the view
point of the liquid-solid transition of
hard spheres.  This will certainly help one to compare the
configurational statistics and other thermodynamic properties 
of vibrating beds with those equilibrium properties of hard sphere
systems.
\vskip 0.3 true cm
{\it Transition temperature and the thickness of boundary layers}:
Consider a collection of elastic hard spheres of mass m and diameter D, 
confined
in a two dimensional $(x,z)$ 
container with an open boundary at the top.  Gravity acts along 
the negative z direction.
The system is in contact
with the thermal reservoir at a temperature $T$ in such a way 
that the average kinetic energy
of each hard sphere, $T=m<v_x^2+v_z^2>/2$ with $<..>$ being the 
configurational average. 
We now start from
$T=0$, at which point 
all the particles are essentially in a solid regime, and the density 
profile is simply 
a rectangle.
If we gradually increase the temperature, fluidization starts from the surface,
and the boundary layers appear.
One may estimate the thickness of the
boundary layer, $h$,  by a simple energy balance between 
the kinetic and potential energy:
$ mgDh \approx \frac{1}{2}m<v^2> \approx T$.  From this, one may
obtain the size of the solid-like regime,
or equivalently its dimensionless height, say $\zeta_F(T)$:
$$ \zeta_F(T)= \mu-h = \mu -\frac{T}{mgD} \eqno (1)$$
where $\mu$ is number of layers
of the original rectangle.  
Eq.(1) predicts the existence of
a critical temperature, $T_c$, at which point
a phase transition from a one phase(solid) to a two
phase regime(liquid-solid) occurs.  By setting, $\zeta_F(T_c)=0$, we find
the mean field result:
$ T_c^{M.F} = \mu mgD$.
Since the boundary layer exists only when both phases coexist, $T_c$
must be the temperature at which point the system becomes fully
fluidized.  
One may equally define the critical temperature as a point at which
the density at the bottom layer, $\phi_o$,
becomes the closed packed density $\phi_c$, i.e: $\phi_o(T_c)=\phi_c$.
We now rewrite eq. (1) in terms of the critical temperature, and recast
the size of the solid
region, in term of $T/T_c$, as
$$\zeta_F(T)/\mu=(1-\frac{T}{T_c}) \eqno (2)$$
\vskip 0.3 true cm
A more precise estimate of the
transition temperature was given 
in ref.[6] within the framework
of Enskog theory [9]. In particular, the following expression for
the density profile, $\phi(\zeta)$,
was obtained as a function of the dimensionless variable $z/D$:
$$-\beta(\zeta-\bar\mu) = ln\phi + c_1\phi + c_2log(1-\alpha\phi)
+ c_3/(1-\alpha\phi) + c_4/(1-\alpha\phi)^2\label{(15)}\eqno (3a)$$
with the constant $\beta\bar\mu$ given by:
$$\beta\bar\mu = ln\phi_o + c_1\phi_o + c_2ln(1-\alpha\phi_o)
+c_3/(1-\alpha_o) + c_4/(1-\alpha\phi_o)^2 \equiv f(\phi_o) \eqno (3b)$$ 
where $\beta=mgD/T$, and 
$c_1 = 2\alpha_2/\alpha^2\frac{\pi}{2} \approx 0.0855$,
$c_2 = -\frac{\pi}{2}
(\alpha_1 - 2\alpha_2/\alpha)/\alpha^2\approx -0.710$
$c_3=-c_2$,$c_4= \frac{\pi}{2}(1-\alpha_1/\alpha + \alpha_2/\alpha^2/)
\alpha \approx 1.278$.
Note that the relation between the volume fraction $\nu$ and 
$\phi$ is given by:  $\nu=\pi(D/2)^2N/V = \pi\phi/4$.
If one integrates the density profile, and imposes the sum rule,
$\int_0^{\infty} \phi(\zeta)d\zeta=\phi_o\mu$, then one finds 
that this sum rule breaks down at the temperature $T_c$,
$$T_c = mgD\mu\phi_o/\mu_o \eqno (4)$$
The departure from the mean field theory is
the appearance of a factor $\phi_o/\mu_o$ in (4), where $\phi_o=(4/\pi)(
\pi/2 \sqrt{3})=1.154700538..$ and $\mu_o=111.52274..$.  Eq. (2)
remains unchanged.
We now present MD data to test Eq.(2) and (4).
\vskip 0.3 true cm
{\it Molecular Dynamics simulations of gravity induced liquid-solid 
transition}:
We have used the Event Driven(ED) Molecular Dynamics code, and
refer the readers to references [10] for
details of the algorithm regarding the collision dynamics
that take into account the rotation of hard spheres,
and a way to handle the inelastic collapse.
The thermal reservoir of our system was modeled using white noise driving
[11], which kicks each particle so that the average kinetic
temperature of each particle is the same as that of
the reservoir, and hence, the kinetic temperature of the
system.  Note that we are {\it not} driving the system by connecting
the bottom wall to the temperature reservoir, which was often used as
a model for a vibrating bed.

We present in Fig.1 a typical configuration below the
transition temperature ($T<T_c$), at which
about 17 layers of
particles condense and form a crystal near the bottom(Fig.1a).
More precisely, the particles first form
a loose hexagonal crystal 
and progressively evolve into a compact
hexagonal lattice structure.  The solid line in the density profile(Fig.1b)
is the Enskog profile given by (3a), which was {\it shifted} to fit the
data beyond the crysal regime.  We point out here
that (i) this shift is {\it not} an arbitrary parameter, but
should be uniquely chosen to fit the data, (ii) this shift in fact
determines the {\it measured} size of the solid by simulations.
The density in the solid regime is then 
fit by a straight line as shown in the figure.  
The oscillations in the solid regime are real, but it is simply the finite
size effect, i.e, the hexagonal packing in a finite lattice has two more
particles in alternative layers.  This oscillation must disappear in the
thermodynamic limit.

The critical temperature $T_c$ is determined as
the temperature at which point a {\it compact} hexagonal crystal is
formed from the bottom layer, beyond which point, 
the density at the bottom layer remains constant at $\phi_o=1.15$, and
this hexagonal structure is permanelty retained. We point out
that a loosely hexagonal crystal forms at a temperature, $T_c'$, which is
somewhat
larger than $T_c$.  Between $T_c$ and $T_c'$, particles
squeeze themselves, expelling holes, and progressively forming a compact
hexagonal crystal. Note that a few
vacancies created
during this crystallization do not anneal but stay in the system (Fig.1a).
Now, in order to carry out the quantitative analysis of the formation of
a crystal beyond the transition temperature,
we have
{\it measured} the size of the solid as mentioned above, namely by
shifting the Enskog profile (i.e. Fig.1b),
and  plotted it at different temperature $T < T_c$
as a function of the scaled variable $T/T_c$
for 1000 particles of $m=2.090\bullet 10^{-6}$,
$D=0.001m$ and $\mu=20$.  The solid line in Fig.2a
is the prediction Eq.(2).  The excellent agreement between the theory and
simulations is a confirmation of (i) the
existence of the gravity induced liquid-solid
transition of hard spheres, and (ii) the validity of the suggested
mechanism of this transition via the disappearance of particles from
the liquid 
and their settlement into the solid regime as predicted by Enskog theory [6].

Next, we present our new analysis of the vibrating bed from the
view point of the liquid-solid transition discussed above.
It has been fairly well established that the configurational
statistics of the vibrating beds seem identical to the equilibrium
statistics of a molecular gas at an equal packing fraction [8], yet the
relation between the vibrational strength, $\Gamma$,
and the corresponding equilibrium kinetic
temperature has remained largely undetermined.  There has been a
previous attempt to relate $\Gamma$ to
the Fermi temperature [12], which is not the same as the kinetic
temperature, but essentially the compactivity [13].
In the present work, we will establish a specific
relation between the vibration strength and
the kinetic temperature, and test its validity via
simulations.

At a low vibration strength, experimental
data [7] seems to clearly indicate two distincitve 
regimes: solid regime near the bottom where there are very little particle 
movements, and the liquid regime near the surface where particles are
dynamically active exchanging their positions via collisions.  Hence,
the system presented in re.f [7] is {\it below} the liquid
-solid transition temperature.  We
will determine both the transition temperature, $T_c$, and the effective
temperature of the system, and then measure the
size of the solid region and compare it with the prediction made by
Eq.(2).  The control parameters are given in ref.[7], namely the particle
diameter $D=2.99 mm$, and the dimensionless initial layer thicnkness
$\mu=10.2$, from which we determine the normalized critical temerature
of the vibrating bed
$T_c/mg=\mu D\phi_o/\mu_o= 0.607 mm$.  The effective temerature of the
system is then determined by fitting the tail region and shifting the
Enskog profile, by $\zeta_o$.  We find, $T/mg=0.36mm$, 
and $\zeta_o=4.41\qquad layers$ 
from which we measure the size of the solid
as $z_o=\phi_o\zeta_o/D = 12 mm\approx 4.0 \qquad layers$, while 
the predicted dimensionless height of the solid region, $\zeta_F$, is:
$\zeta_F = \mu (1- T/T_c) \approx 4.15 \qquad\qquad layers$.
The previous fitting of the density profile by
the Fermi profile was also satisfactory, but was found to
be most difficult near the rounded
 region, which the Enskog profile fits quite well.
One advantage of the present method of analyzing
the configurational statistics of the vibrating bed might be that the global
{\it kinetic} temperature can now be associated with the vibrating bed, 
and hence comparison can be made between the
experimentally determined configurational statistics of the vibrating bed
and those of the hard spheres in thermal contact with the heat reservoir.
The specific relation between the two can be obtained by comparing the
thermal expansion of the hard spheres and the kinetic expansion of the 
vibrating bed.  The thermal expansion is simply the increase in the center
of mass $\Delta z(T)$, 
which can be computed by the Enskog profile near the
tail, and the solid rectangle.
We find:
$$\Delta z(T)=\frac{D\mu}{2} (\frac{2|\Lambda_1|\phi_o}{\mu_o^2}-1)
(\frac{T}{T_c})^2 \eqno (5)$$
where the constant $|\Lambda_1|=\phi_o\int_o^1[f(\rho\phi_o)-f(\phi_o)][\rho
\phi_o f'(\rho\phi_o)d\rho] = 5503.531806$ with $f(x)$ given in (3a).  Note
that the correction is second order in $T$.  Let $H_o(\Gamma)$ be
the single ball jump height on the surface [16].  Then, by equating 
$\Delta z(T)$ and $H_o(\Gamma)g/\omega^2$, we find the desired relation:

$$\frac{T}{T_c} = \sqrt{\frac{2H_o(\Gamma)g}{D\omega^2\mu^2}
(\frac{\mu_o^2}{2|\Lambda_1|\phi_o-\mu_o^2})} \eqno (6)$$
where $\omega$ is the vibration frequency.  Putting all the values, eq.(6)
predicts $T/T_c=0.663$, which is close to the measured value of
$0.593$ above.

In conclusion, two points are in order.
First, we have demonstrated in this Letter that
the point at which the Enskog
description of hard spheres fails indeed signals the liquid-solid transition,
and such a failure arises via the breakdown in
the particle conservation. The
missing particles form a condensate at the bottom, which
essentially
determine the
fraction of particles in the solid regime,
and in turn the thickness of boundary layers.  Since only a fraction
of grains are mobilized under shear [14], and avalanches and
many interesting dynamics occur in these thin boundary layers [2],
such a determination should be of technological importance.
Second, since 
Enskog theory is a truncation of BBGKY [15] hierarchy at the third order,
the existence of gravity induced liquid-solid transitions of
hard spheres must have some interesting
consequences to higher order
kinetic theory, in particular with regard to the dynamic
behaviors.  Unlike particles in the liquid regime,
those particles in the solid regime 
are largely confined in cages and fluctuate around fixed
positions.  Their motions resemble the
lattice vibrations rather
than binary collisions, and it may be a little peculiar,
albeite not unphysical, to attempt to
describe the lattice vibrations by the kinetic theory. If so, 
such a description must include much more than binary collisions.
Hence, it is not
unphysical to see that these particles disappear from the kinetic equation 
{\it at
the level} of the Enskog approximation.  
However, as discussed in the beginning and
demonstrated in this paper, this gravity induced liquid-solid 
transition is not a peculiar phenomenon
associated with Enskog equation, but rather an intrinsic transition
inherent in a system where an excluded volume interaction is dominant.
The formation of a solid at the bottom is the appearance of a massive
occupied low energy state due to the Pauli exclusion principle.
Therefore, the breakdown in the sum rule,
the necessary shift of the density profile due to the
formation, and its upward spread of the closed packed regime 
{\it should} persist because the Pauli exclusion principle is in action
in real space,
even if one may use different approximations [16-18]  or may try
a different form for the pressure, such as the form suggested by
Percus-Yevick [19], and/or in higher order truncation.
It only disappears in the limit when the
closed volume packing density, $\nu$ becomes one,
which is possible only in the case of an ideal
Appolonian packing [20].
Finally, we point out that the presence of dissipation does not alter
the condensation picture at all [21], {\it if} the velocity distribution
remains Gaussian.  Recent experiments [22] have demonstrated the non-Gaussian
nature of the velocity distribution, but if the dissipation is small, which is 
the case for the simulations carried out in this work, the deviation from
Gaussian should be small.
\vskip 1.0 true cm
\noindent {\bf Acknowledgement}
We wish to thank Stefan Luding for providing us with his MD code and many
helpful discussions over the course of this work.

\newpage
\noindent {\bf References}

\noindent [1] J-P Hansen, McDonald, Theory of Simple Liquids, 
2nd Editions, Academic Press, London, 1986.

\noindent [2]  H. Jaeger, S.R. Nagel, and R. P. Behringer, Rev. Mod. Phys. 68,
 1259 (1996); See, also 'Granular Gases', Edited by S. Luding, Poschel,
H. Herrmann, Springer-Verlag (2000).

\noindent [3] See for example, Y.B. Kim, ``Strongly Correlated Electrons in
the Quantum Hall Regime,'' Newsletter, The Association of
Korean Physicists, Vol.21. 29 (2000) and references threin.

\noindent [4] H. Frohlich, International J. Quan. Chem. Vol. II. 641 (1968).

\noindent [5] H. Haken, ``Synergetics: nonequilibria, phase transitions and
self organization. Naturwisse nschaften, Vol. 68, No.6. 293 (1981).

\noindent [6] D. C. Hong, Physica A 271, 192 (1999).

\noindent [7] E. Clement and J. Rajchenbach, Euro. Phys. Lett. 16, 133 (1991).

\noindent [8] S. Warr and J. P. Hansen, Euro. Phys. Lett. {\bf 36}, 589
(1996).

\noindent [9] D. Enskog and K. Sven, Vetenskapsaked Handl. 63, 4 (1922).
 S. Chapman and T. G. Cowling, The Mathematical Theory of
Nonuniform Gases, Cambridge, London, 1970.

\noindent [10] B. D. Lubachevsky, J. Comp. Phys. 94, 255(1991).
S. Luding, Phys. Rev. E. 52, 4442 (1995).
S. Luding and S. McNamara, Granular Matter,1(3), 113 (1998).

\noindent [11] D. M. Williams and F.C. MacKintosh, Phys. Rev. E 57, R9 (1996).

\noindent [12] H. Hayakawa and D. C. Hong, Phys. Rev. Lett. {\bf 78}, 2764 
(1997).

\noindent [13] S. F. Edwards and R. B. S. Oakeshot, 
Physica A {\bf 157}, 1080 (1989); A. Mehta and S.F. Edwards, Physica
 A {\bf 168}, 714 (1990).

\noindent [14] D.M. Haynes, D. Inman, J. Fluid. Mech. {\bf 150}, 357 (1985).
See also: L.E. Silbert et al, APS March Meeting Bulletin, P25.007 (2000).

\noindent [15] J.G. Kirkwood, J. Chem. Phys. {\bf 7}, 
911 (1939).  N. N. Bogolyubov. J. Phys. USSR {\bf 10}, 257 (1946).
M. Born and M. S. Green, ``A General Kinetic Theory of Liquids.''
Cambridge University Press, Cambridge (1949).

\noindent [16] G. Rascon, L. Mederos, and G. Navascues, Phys. Rev. Lett.
{\bf 77}, 2249 (1996). T. P. C. van Noije and M. H. Ernest, Granular Matter
{\bf 1}, 57 (1998).  B. Doliwa and A. Heuer, Phys. Rev. Lett.
{\bf 80}, 4915 (1988). S. Torquato, Phys. Rev. E {\bf 51}, 3170 (1995).
J. J. Brey, J. W. Dufty, C.S. Kim, and A. Santos, Phys. Rev. 
E {\bf 58}, 4638 (1998).

\noindent [17] A. Santos, S. B. Yuste, and M.L.D. Haro, Mol. Phys. {\bf 96}, 
1 (1999).  P. Richard, L. Oger, J.-P. Troadec, and 
A. Gervois, Phys. Rev. E
{\bf 60}, 4551 (1999); P. Sunthar and V. Kumaran, Phys. Rev. E {\bf 60},
1951 (1999).  V. Kumaran, Phys. Rev. E {\bf 59}, 4188 (1999).

\noindent [18] S. Luding, in {\it Granular Gases,} edited by
T. Poschel and S. Luding (Springer Verlag, Berlin, 2000), and preprint (2000).

\noindent [19] J.K. Percus and G.J. Yevick, Phys. Rev. {\bf 110}, 1 (1958).

\noindent [20] For Appolonian packing, see B. Mandelbrot, ``The Fractal
Geometry of nature,''(W. H. Freeman and Company, New York, 1982).

\noindent [21] J.S. Olafsen and J.S. Urbach,
Phys. Rev. E {\bf 60}, R2468 (1999).
W. Losert, D. Cooper, D. Kudrolli, and J.P. Gollub, Chaos, {\bf 9}, 682 (1999).

\noindent [22] D. C. Hong, Fermi Statistics and Condensation, To appear
in 'Granular Gases,' edited by S. Luding and T. Poschel
Springer-Verlag (2000).

\newpage
\noindent {\bf Figure Captions}

\noindent Fig.~1.
(a) Snap shot at $T<T_c$, where about 17 layers form a crystal. 
(b) The fitting of the density profile is the combination of the Enskog profile
(Eq.3) and the rectangle( straight line).

\noindent Fig.~2.  The fraction of the hard spheres in the condensed regime
as a function of $T/T_c$ with 
$N=1000$, $\mu=20$, $g=981cm/sec^2$, and $m=1.047\bullet 10^{-6}$.
(square).  The data points are obtained by uniquely determining the
shifting position of the Enskog profile, and the solid line is the
prediction Eq.(2).

\noindent Fig.~3. Experimental density profile of the granular materials
in a vibrating bed (ref.5).  The fit was by the Enskog profile near the 
surface, and the rectangle below $\zeta_F$.  


\newpage
\thispagestyle{empty}
\centerline{\hbox{
\psfig{figure=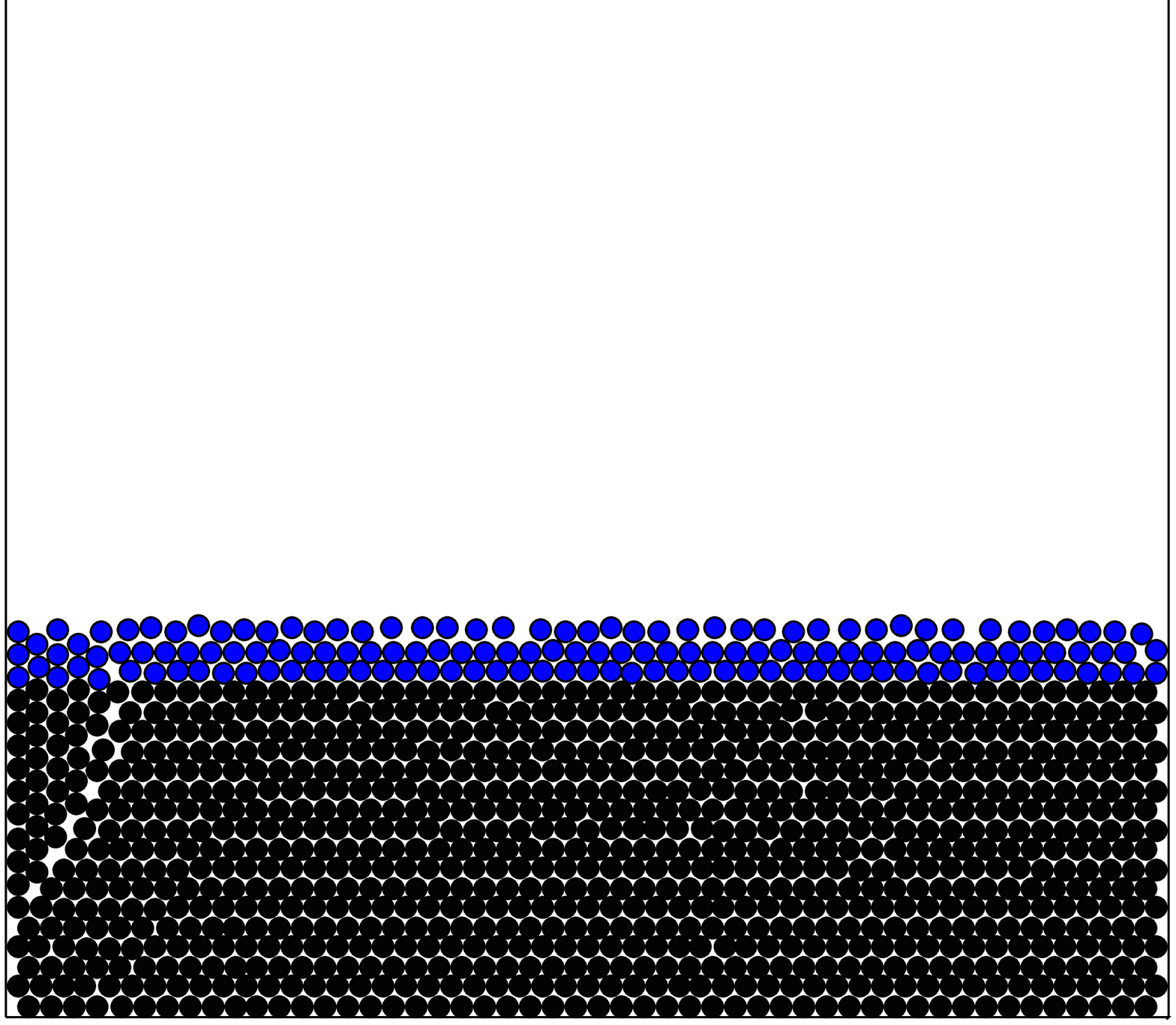}
}}
\newpage
\thispagestyle{empty}
\centerline{\hbox{
\psfig{figure=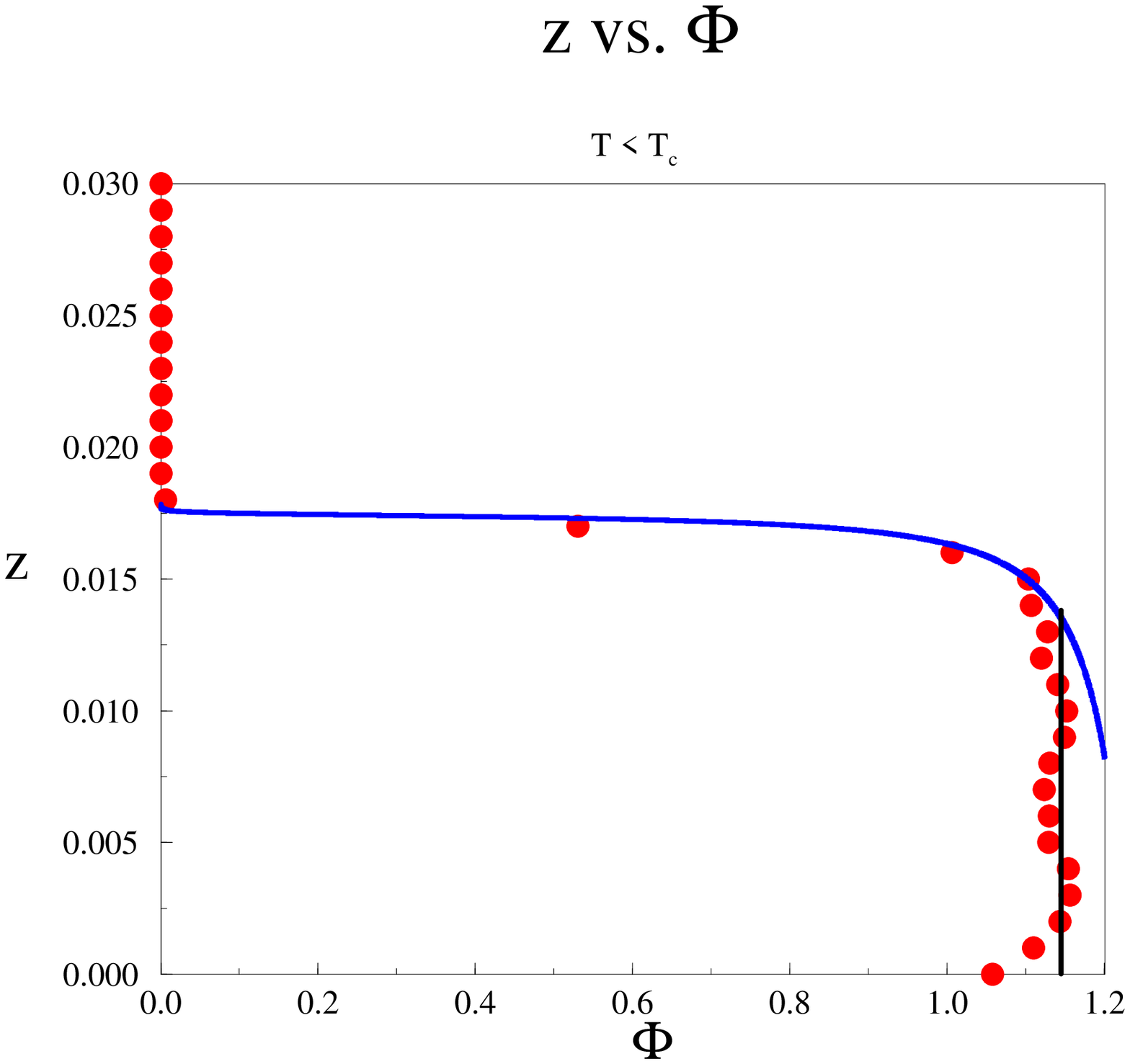}
}}
\newpage
\thispagestyle{empty}
\centerline{\hbox{
\psfig{figure=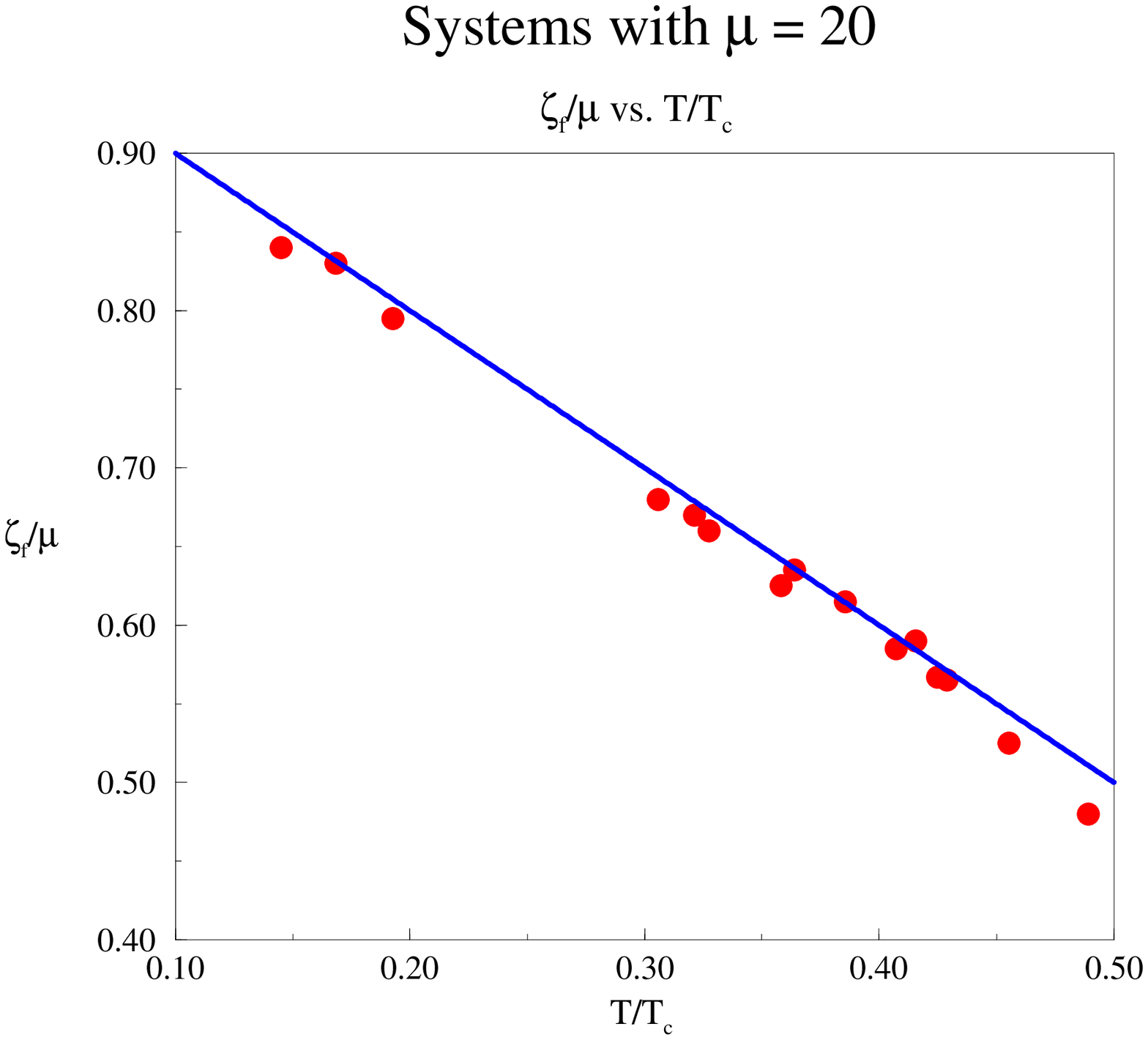}
}}
\newpage
\thispagestyle{empty}
\centerline{\hbox{
\psfig{figure=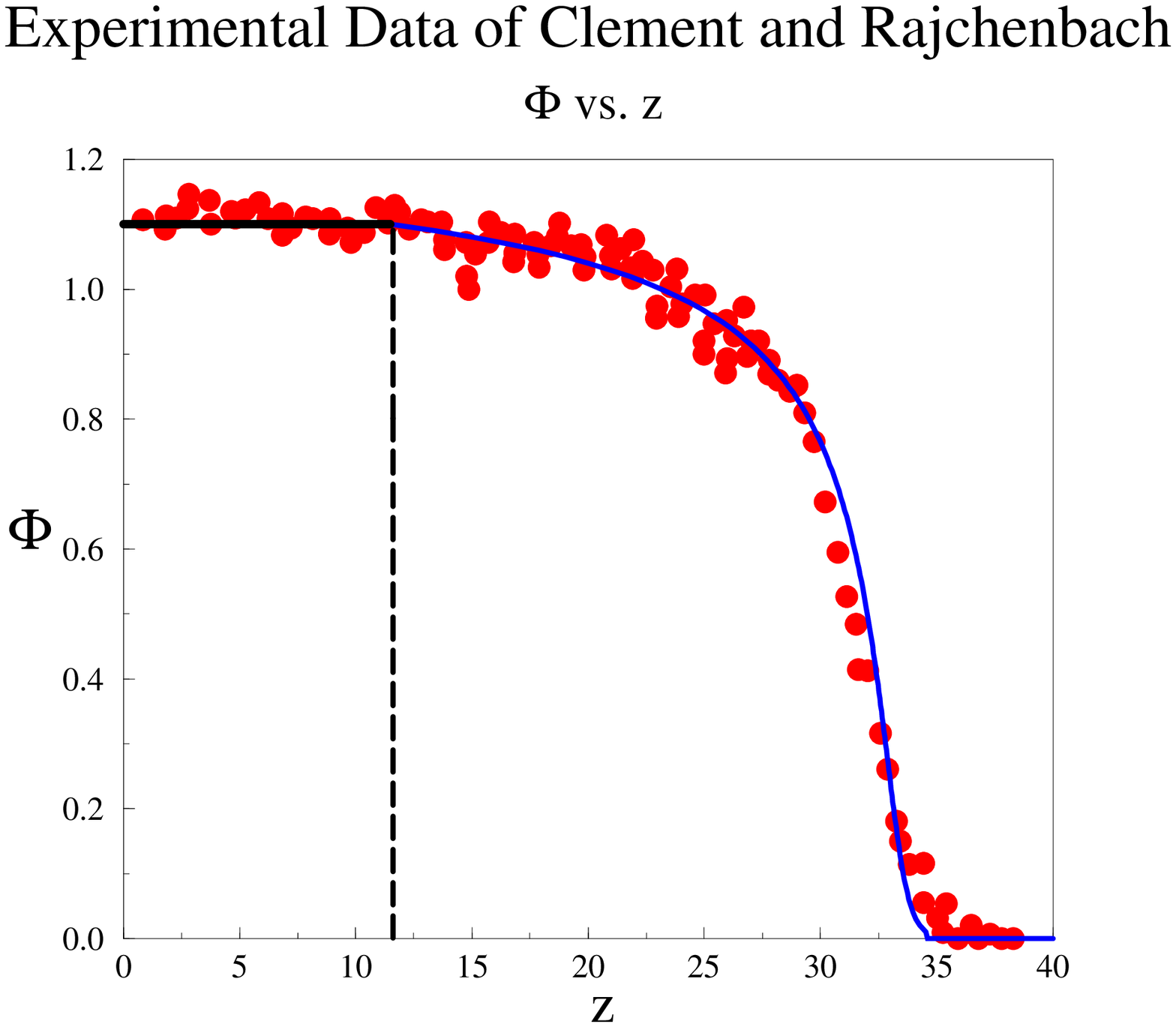}
}}

\end{document}